\documentclass[runningheads]{llncs}
\usepackage{graphicx}
\usepackage{url}
\usepackage{algpseudocode}
\usepackage[utf8]{inputenc} 
\usepackage[T1]{fontenc}    
\usepackage{hyperref}       
\usepackage{url}            
\usepackage{booktabs,comment}       
\usepackage{amsfonts}       
\usepackage{nicefrac}       
\usepackage{microtype}      
\usepackage{xcolor}         
\usepackage{amsmath}
\usepackage{caption}
\usepackage{floatrow}
\usepackage{array, caption, floatrow, tabularx, makecell, booktabs}%
\usepackage{subcaption}
\usepackage[misc,geometry]{ifsym}
\usepackage{graphicx}
\usepackage{placeins}
\usepackage{grffile}

\begin{document}
\title{Cooperative Multi-Agent Search on Endogenously-Changing Fitness Landscapes}
%
\titlerunning{Cooperative Search in Complex Business Domains}
%

\author{Chin Woei Lim\inst{1} \and
Richard Allmendinger\Letter\inst{1}\orcidID{0000-0003-1236-3143} \and
Joshua~Knowles\inst{2}\orcidID{0000-0001-8112-6112} \and Ayesha Alhosani\inst{1} \and Mercedes~Bleda\inst{1}\orcidID{0000-0002-0197-0936}}
\authorrunning{Lim et al.}
%
\institute{The University of Manchester, Alliance Manchester Business School, Manchester M15 6PB, UK\\ \email{brandonwoeilc@gmail.com, \{richard.allmendinger,ayesha.alhosani,mercedes.bleda\}@manchester.ac.uk}\and
Schlumberger Cambridge Research, Cambridge CB3 0EL, UK}
\maketitle              
\begin{abstract}
We use a multi-agent system to model how agents (representing firms) may collaborate and adapt in a business `landscape' where some, more influential, firms are given the power to shape the landscape of other firms. The landscapes we study are based on the well-known NK model of Kauffman, with the addition of `shapers', firms that can change the landscape's features for themselves and all other players. Our work investigates how firms that are additionally endowed with cognitive and experiential search, and the ability to form collaborations with other firms, can use these capabilities to adapt more quickly and adeptly. We find that, in a collaborative group, firms must still have a mind of their own and resist direct mimicry of stronger partners to attain better heights collectively. Larger groups and groups with more influential members generally do better, so targeted intelligent cooperation is beneficial. These conclusions are tentative, and our results show a sensitivity to landscape ruggedness and ``malleability'' (i.e. the capacity of the landscape to be changed by the shaper firms). Overall, our work demonstrates the potential of computer science, evolution, and machine learning to contribute to business strategy in these complex environments. 

\keywords{Cooperative learning \and NK models \and Endogenously-changing landscape \and Shaping \and Searching \and Adaptation}
\end{abstract}

\section{Introduction}

Most non-trivial social systems are inherently challenging to gauge due to the potential complexity arising from interactions at both individual and collective levels~\cite{forrester1971counterintuitive}. Especially in the business context, the mechanics of interaction between competing firms (agents) are often based on rather coarse simplifications and an incomplete understanding of the business landscape. The sophistry embedded within the interplays of businesses, difficult to appreciate from the outside, produces counterintuitive resultant behaviours~\cite{forrester1970urban}. 

Firms compete by developing new strategies, technologies or business models all of which involve solving complex problems and making a high number of interdependent choices. To solve these problems, managers need to search their firms' business landscapes and find a combination of these choices that allows them to outperform their competitors. Bounded rational managers cannot easily identify the optimal combination, and tend to engage in sequential search processes~\cite{SimonandMarch1958} and, via trial and error, learn and find what combinations are possible and perform well. Effective search for well-performing solutions in a business landscape is thus a source of competitive advantage for companies. 

Conceptually, a business landscape dictates the effectiveness of a firm's search strategy by assigning them a fitness, which typically represents the level of return or performance. The active revision of a firm's choices is crucial in maintaining its competitive advantage, growth and profitability when competing against other firms on a business landscape. Such revisions are normally in the form of research and development of any aspect of a firm in order to find better choices (strategies, methods, and/or products), leading it towards a better path, and to higher local peaks on the landscape. Generalising, firms improve their performance by adapting to the business landscape within which they operate. However, actual business landscapes are dynamic, and they tend to change not only exogenously as a result of external factors (changes in government policies and regulations, in demographic and social trends, etc) but also due to the behaviour and strategies of the firms competing within them. Firms simply do not limit themselves to only adapting and accepting the state of their current environment as it is. Capable firms might be able to \emph{shape} the business landscape to their advantage (in addition to \emph{search} the landscape)~\cite{felin2014economic,uzunca2018sharing}. A quintessential example of this phenomenon was when Apple introduced the iPhone and swiftly shook the environment in its favour, demolishing Nokia, which was the incumbent cell-phone market leader at that time.

Management research has used the NK model~\cite{ganco2009performance,kauffman1993origins} introduced in the management literature by~\cite{Levinthal997} to build simulation models to represent business landscapes, and study different factors that influence the effectiveness of companies' search processes (see~\cite{Baumann2019} for a review). Despite the usefulness of these models, most of them consider that the business landscape within which companies compete does not change or it changes in an exogenous manner, i.e. due to factors external to the companies. They thus do not account for the influence that endogenous changes rooted in firms' behaviour have in the business landscape, and in turn in the performance of firms within it. The first simulation model that has analysed companies search effectiveness when business landscapes change endogenously was proposed by Gavetti et al in 2017~\cite{gavetti2017searching}. The authors extend the NK model to consider two types of firms (agents): agents that search a landscape only (referred to as \emph{searchers}) and agents that can both search and shape the landscape (\emph{shapers}). Consequently, searching firms need to adapt (search) on a landscape that is being shaped (changed) by the shaping firms. In other words, shapers have the power to change a landscape endogenously, while searchers perceive these changes as exogenous. Since all agents (shapers and searchers) search the same landscape, a change in the landscape (caused by a shaper) affects all agents. The study of Gavetti et al.~\cite{gavetti2017searching} focused on studying how the impact of different levels of landscape ruggedness and complexity, and the proportion of shapers vs searchers operating in it, affect the performance of both types of firms. In real contexts, firms do not only compete within a business landscape but in many cases competing firms try to improve their performance by cooperating, i.e. via coopetition. Coopetition is the act of cooperation between competing companies. Businesses that engage in both competition and cooperation are said to be in coopetition. This paper extends Gavetti et al~\cite{gavetti2017searching} by allowing firms to cooperate and analyses how cooperation influence firms performance in endogenously changing business landscapes. This is achieved by incorporating cognitive and experiential search into the adaptation process.

The next section details the traditional NK model and the adapted version of that model by Gavetti et al.~\cite{gavetti2017searching}  (also referred to NKZE model), and explains the search rules. Section~\ref{StealCoop} proposes a cooperative approach with learning, and Section~\ref{ExStudy} provides details about the experimental study and then analyzes the proposed approach for different configurations of the simulated changing (and competitive) business environment. Finally, Section~\ref{conclusion} concludes the paper, and discusses limitations of the work and areas of future research. 

\section{Preliminaries}\label{Prelim}
\subsection{Kauffman's NK(C) Model}
The NK model of Kauffman~\cite{kauffman1993origins} is a mathematical model of a tunably rugged fitness landscape. 
The ruggedness is encapsulated by the size of the landscape and the number of local optima, which are controlled by the parameters, $N$ and $K$, respectively. Formally, in an NK model, the fitness $f(\vec{x})$ of an agent (firm) at location $\vec{g}=(g_1,\dots,g_N)$, $g_i\in \{0,1\}$, on the landscape can be defined as
\begin{equation}
f(\vec{g}) = \frac{1}{N}\sum_{i=1}^{N}f_i(g_i, g_{i_1},\dots,g_{i_K}),
\end{equation}
where $g_i$ is the $i$th (binary) decision variable, and the fitness contribution $f_i$ of the $i$th variable depends on its own value, $g_i$, and $K$ other variable values, $g_{i_1},\dots,g_{i_K}$. The parameter $K$ has a range of 0 $\leq$ $K$ $\leq$ $N-1$ that determines how many other $K$ different  $g_i$'s will be affecting each $g_i$ when computing fitness. The relationships between $g_i$'s are determined randomly and recorded in an \emph{interaction matrix} that shall be left unchanged. The function $f_i:\{0,1\}^{K+1}\rightarrow \mathbf{R}$ assigns a value drawn from the uniform distribution in the range [0,1] to each of its $2^{K+1}$ inputs. The values ${i_1},\dots,i_K$ are chosen randomly (without replacement) from $\{1,\dots,N\}$. Increasing the parameter $K$ results in more variables interacting with each other, and hence a more rugged (epistatic) landscape. The two extreme cases, $K=0$ and $K=N-1$, refer to the scenarios where the fitness contributions $f_i$ depend only on $g_i$ (i.e. each $f_i$ can be optimized independently) and all variables, $g_1,\dots,g_N$, respectively (maximum ruggedness). 


Taking an arbitrary firm with a search policy string of $\vec{g} = (011101)$, we can calculate the fitness contribution ($f_i$) of  $g_1$ by forming a temporary string with the  $g_i$'s that are related to itself by referring to the interaction matrix. Let us assume that, in this example, $g_1$ was initially and randomly determined to be related to $g_2, g_4$ and $g_5$. Since $g_1=0$, $g_2=1$, $g_4=1$ and $g_5=0$, the string formed shall be $(0110)$. The fitness contribution can then be extracted from the fitness matrix by taking the value from $6^{th}$ row (0110 in decimal), and the $i^{th}$ column ($1^{st}$ column in this case). Understandably, the fitness contributions of subsequent  $g_i$'s are calculated similarly.

Kauffmann later extended the NK model to introduce coupled landscapes (a.k.a NKC model)~\cite{kauffman1993origins}, which allows multiple species to exist on different landscapes, and interact through a phenomena of niche construction.

\subsection{Gavetti et al.'s NKZE Model}
The conventional NK model allows firms to continually adapt on a fixed landscape until they reach some local or global optima. Additionally, the action of any firm has no consequence on other competing firms. However, in a realistic and dynamic business environment, the introduction of disruptive technologies and concepts can often drastically restructure the business landscape, thereby needing competing firms to re-strategize towards a new goal or face obsoleteness. 

Gavetti et al.~\cite{gavetti2017searching} introduced the concept of shapers, which have the ability to modify the business context to their own advantage on top of the standard agents (hereinafter known as searchers) in the baseline NK model. They then studied the effects of different levels of shaping on the performance of shapers themselves and on searchers. Unsurprisingly, shaper firms and the level of shaping have great effects on the performance of their competitors as landscape restructuring always tend to undermine competitor performance. However, a high level of shaping coupled with a great number of shapers were found to be highly non-beneficial for shapers and searchers alike, as constant landscape restructuring changes the objective too fast and too much, thus rendering local search obsolete and causing massive performance instabilities.

The key feature of shapers is their ability to influence the business context (hopefully) to its own advantage, and as a side-product, alter the fitness of their competitors. To achieve this, Gavetti et al.~\cite{gavetti2017searching} extends the NK model with an additional $Z$ (binary) decision variables, $\vec{e}=(e_1,\dots,e_Z), e_i\in\{0,1\}$. Here, $\vec{e}$ is referred to as the shape policy string and is globally shared by all firms, differently to the search policy string, $\vec{g}$, which is controlled by each firm (agent) independently. Similarly to the parameter $K$ in the NK model, Gavetti et al.~\cite{gavetti2017searching} use a parameter $E$ to interlink the shape and search policy strings: each $g_i$ is related to $E$ randomly sampled $e_i$'s or $e_{i_1}\dots,e_{i_E}$. Such relationships are also recorded in an interaction matrix that shall be kept constant throughout a run. Notably, $E$ has a range of 0 $\leq$ $E$ $\leq$ $Z$.

Accordingly (for the added dimensions), Gavetti et al.~\cite{gavetti2017searching} update the fitness assignment function to $f_i:\{0,1\}^{K+1+E}\rightarrow \mathbf{R}$, which assigns a value drawn from the uniform distribution in the range [0,1] to $2^{K+1+E}$ inputs. 
In practice, the fitness contributions are stored in a matrix of size $2^{K+1+E}\times N$, and the interactions are stored in a matrix of size $N\times K+1+E$. Now, the evaluation of a firm's fitness depends on $E$ too. Taking the search policy string for an arbitrary firm to be $\vec{g}=(011011)$, global shape policy string to be $\vec{e}=(101000)$, $K=3$ and $E=3$, an interaction matrix was generated randomly, resulting in $g_{1}$ being related to $g_{2}, g_{4}, g_{5}, e_{2}, e_{3}$ and $e_{6}$. The fitness of contribution of $g_{1}$ ($f_{1}$) will be determined by firstly forming a temporary string in the order of $(g_{1}g_{2}g_{4}g_{5}e_{2}e_{3}e_{6})$, making $(0101010)$. Similarly, $f_{1}$ will be taken as the $42$nd ($0101010$ in decimal) row and the $1$st column ($ith$ column) of the fitness contribution matrix.

Gavetti et al.~\cite{gavetti2017searching} use the following approach to tackle the NKZE Model: At the beginning of the simulation, a predetermined amount of firms will be turned into shapers based on a shaper proportion ($\beta$) parameter. Firms are re-ordered randomly at the beginning of each iteration. More specifically, all firms are allowed to make an action in accordance with the randomly determined order in each iteration. Thus, the number of actions within an iteration would be equal to that of the firm (agent) population. In terms of action, each (shaping and searching) firm is allowed to make one adaption move. A searching firm flips one randomly selected search policy bit (keeping the shaping policy unaltered) and if the resulting policy has a better fitness than the current policy, then the firm retains the new policy; otherwise, the firm will stick with the old policy.  However, when it is the turn of a shaper, it has the choice of either adapt as a searcher would without altering the shape policy string, or randomly mutating a single bit of the shape policy string and evaluate fitness with its original unmutated search policy string. A shaper will then pick either choice that is better, or end its turn without adopting any mutation if both the choices were found to be unfit. 

Intuitively, $E$ also corresponds to the level of shaping and the malleability of the fitness landscape. A higher $E$ means that the globally shared shape policy string has more influence on fitness contributions, and transitively, the fitness landscape itself. Under this condition, the extent of fitness landscape restructuring when a shaper acts on the shape policy string is high. Thus, the fitness landscape is said to be highly malleable at high $E$.

The NKZE model is a variation of Kaufmann’s NKC model. While both models allow agents to dynamically change the environment, there are critical differences: in the NKC model (i)~each species operate on a separate landscape, (ii)~all species have the ability to change the landscape, and (iii)~each species is represented by one agent only. Consequently, the two models are designed to simulate different (simplified) business environments. 


\section{Stealthy and Cooperative Learning}\label{StealCoop}
The transient environment of the NKZE model caused traditional myopic ``hill-climbing'' adaptation to underperform. Additionally, such myopic practices hardly capture the rationality of realistic firms. Motivated by this weakness, we extend the methodology to tackle the NKZE model to provide firms with the ability to learn from one another (stealthily or cooperatively), potentially allowing firms to adapt more quickly and adeptly to changing environments. 

To implement such ideas, a strategy of exploiting multi-agent search in NKZE with population-based optimisation techniques, specifically particle swarm optimisation (PSO)~\cite{Engelbrecht2015PSO} and explicit direct memory genetic algorithm~\cite{Yang2007MemoryGA}, was implemented. This was done by (i)~allowing firms to quickly adapt to the environment by looking towards excellent firms during the exploration phase following concepts inspired by PSO (similar to neighborhood search~\cite{wang2013diversity}) and (ii)~preserving good solutions in a memory and exploiting them at the end of exploration. We will refer to this strategy as stealthy global learning (StealthL). StealthL operates in an idealistic environment where intel regarding the strategies (and success level) of competing firms is always freely and readily available without limitation (i.e., globally). However, such limitation does exist and is inherent to the nature of competition. Additionally, the NKZE and StealthL model do not share similar dynamics, as the former had a single-mutation restriction whereas the latter allowed for very rapid adaptation by mutating multiple elements within policies of firms. The dynamics of NKZE is more realistic as a change in a firm's policy takes time and is limited by resources. A complete or near-complete revamp of policies continuously is not affordable.

As a result, the StealthL model was modified to allow firms to form collaboration groups. Swarm intelligence and memory scheme were now restricted within the boundary of these groups, thus limiting the amount of information a firm gets. This new model will be referred to as the structured cooperation (StructC) model. Both StealthL and StructC were compared against the standard adaptation used in the NKZE (hereinafter known as the standard model) in the next section. First, we will describe StealthL and StructC in the next two subsections.

\subsection{Stealthy Global Learning}

Our model of how stealthy learning occurs between firms is based on a simple information-sharing scheme used in the swarm intelligence method, PSO. This is augmented with the use of a memory of past policies, a technique reminiscent of poly-ploid organisms' storage of defunct (inactive) genetic material (chromosomes) that can be resurrected quickly under environmental stress.

\vspace{+1mm}
\noindent \textbf{Swarm Intelligence\quad}To implement swarm intelligence, the search policy string of an arbitrary firm was mutated based on a guiding vector that is unique to each firm~\cite{yang2007learning}. Descriptively, the guiding vector of an arbitrary firm is $\vec{P}=(p_{1},\ldots,p_{N})$ and has a length of $N$, matching that of the search policy string $\vec{g}=(g_{1},\ldots,g_{N})$. Each element $p_{i}$ in the guiding vector represents the probability of its corresponding element ($g_{i}$) in the search policy string to mutate to 1. Naturally, the probability of which $g_{i}$ mutates to 0 is given by $1-p_{i}$. All guiding vectors were randomly initialised with a uniform distribution with [0,1] range at the beginning. At its turn, the firm first learns towards the search policy string of the current global best performing firm $({g}_{max f,t})$ at time $t$ using its guiding vector with a learning rate of $\alpha$ where $0 \leq \alpha \leq 1$:

\[\vec{P}_{t+1}=(1-\alpha) . \vec{P}_{t}+\alpha . \vec{g}_{max f,t}\;.\] 

Subsequently, a string of random variates with length $N$ is generated as $\vec{R}=(r_{1},\ldots,r_{N})$ using a uniform distribution with [0,1] range. Finally, the new search policy string is determined as follows:

\[g_{i}=\left\{\begin{matrix}
1,r_{i}<p_{i}\\ 
0,r_{i}\geq p_{i} \:.\\ 
\end{matrix}\right.   \] 

Note that $p_{i}$ has a range of $0.05 \leq p_{i} \leq 0.95$ to allow for 5\% random mutation after convergence. This new adaptation was designed to  facilitate fast landscape adaptation via guided multiple mutations. The single random mutation of the shape policy string was kept without alteration to preserve the nature of the landscape-shaping dynamics. Finally, the firm chooses whether to adopt the mutated policy strings or to remain unchanged as in the standard NKZE model.  

\vspace{+1mm}
\noindent \textbf{Learning from Experience, a.k.a. Polyploidy\quad}In addition to swarm intelligence, the model was also extended to memorise the aggregated policy (search + shape) of the best performing firms in each iteration~\cite{Yang2007MemoryGA}. We limit the size of the database in which these memories are stored to $\Theta$ agents. To ensure environmental diversity within the database, newly memorized candidate memory should have a unique shape policy string. If the shape policy string of the candidate is already present in a memory, the fitter one will be adopted. At full capacity, replacement can only happen if an environmentally unique candidate was better than the worse performing memory. 
To prevent premature memory exploitation, a parameter $\varepsilon$ representing the probability of not exploring the database was initialised to 1 at the beginning of the model. A decay parameter $\gamma<1$ was then set to reduce $\varepsilon$ at the end of each iteration using $\varepsilon_{t+1}=\varepsilon_{t}.\gamma$.

At the turn of an arbitrary firm, the firm shall only exploit the memory if a random number generated is greater than $\varepsilon$ without undergoing any exploration (searching and/or shaping). The firm then, without hesitation, adopts the best policies of the best performing memory. Note that a searcher can only adopt the search policy string of the best performing memory, whilst a shaper adopts both search and shape policy strings simultaneously. 

\vspace{+1mm}
\noindent \textbf{Relevance to practice\quad}Coopetition emphasizes the mixed-motive nature of relationships in which two or more parties can create value by complementing each other's activity~\cite{Coopetition2007}. In the stealth model, organizations in the landscape are all competing to reach the best fitness, but also cooperating by sharing knowledge with each other. This resembles the scenario of organizations helping each other to reach a new common goal, such as tech giants Microsoft and SpaceX working together to explore space technology. In their collaboration, the organizations work together to provide satellite connectivity between field-deployed assets and cloud resources across the globe to both the public and private sector via SpaceX's Starlink satellite network~\cite{MicrosoftNews}. At the same time, they are competing to dominate niche segments. If this collaboration succeeds, Microsoft and SpaceX will be dominating the space technology market. 

\subsection{Structured Cooperation}
By forming random collaboration groups, firms are now equipped to exchange landscape knowledge amongst their partners, but have no information outside of the group. Thus, (i) a firm can only refer to the best performing policy within its group, and (ii) each group has a separate memory. Mutation-wise, only the element of the search policy string that has the highest probability of mutating can mutate. If multiple were present, one will be selected at random to mutate. 

\vspace{+1mm}
\noindent \textbf{Relevance to practice\quad}Collaboration between companies can occur in numerous ways, one of which is sharing knowledge and expertise. In today's age, expertise and information are considered valuable strategic assets for organizations~\cite{Ye2019Knowledge}. The StructC model mimics sharing knowledge among a closed pre-determined group of companies that falls under the same management/ownership (a.k.a conglomerate). Examples of this type of corporations are, Alphabet LLC who owns Google, DeepMind; Amazon who owns Audible, Amazon Fresh, Ring to list a few. Despite sharing their knowledge and expertise, collaborating companies work towards one goal while maintaining their independence and decision making~\cite{PatzAnalysis2009}. Also, knowledge sharing becomes the natural required action for the company to reduce costs and save time, and improve efficiency~\cite{Jun2019conglomerate}. 

\section{Experimental study}\label{ExStudy}
This section outlines the model and algorithm parameter settings followed by an analysis of results. Table 1 
provides an overview of the default parameter settings for the the three models to be investigated, Standard Model, StealthL and StructC. Parameter $N$, $Z$ and $\beta$ were chosen in accordance to~\cite{gavetti2017searching} for comparison purposes. We simulate a business environment with a population of $M=10$ agents (firms) and a maximum group size of $\omega_{max}=4$. Table 2 
lists the possible group compositions. The default parameters for the other parameters were set based on preliminary experimentation such that robust results are obtained. 


\vspace{+0mm}
\begin{minipage}[c]{0.2\textwidth}
\vspace{+7mm}
\flushleft
 \label{defaultP}
\scalebox{0.7}{
\begin{tabular}{cc}
\toprule
\multicolumn{1}{c}{Parameter}                           & \multicolumn{1}{c}{Default Values} \\ \midrule
\multicolumn{1}{c}{$N$}                                   & \multicolumn{1}{c}{12}             \\ 
\multicolumn{1}{c}{$Z$}                                   & \multicolumn{1}{c}{12}             \\ 
\multicolumn{1}{c}{$M$}                     & \multicolumn{1}{c}{10}             \\ 
\multicolumn{1}{c}{$\beta$}                                  & \multicolumn{1}{c}{50\%}          \\ 
\multicolumn{1}{c}{$\alpha$}               & \multicolumn{1}{c}{0.2}            \\ 
\multicolumn{1}{c}{$p_{i,ceil}$}           & \multicolumn{1}{c}{0.95}           \\ 
\multicolumn{1}{c}{$p_{i,floor}$}          & \multicolumn{1}{c}{0.05}           \\ 
\multicolumn{1}{c}{$\Theta$}                         & \multicolumn{1}{c}{50}             \\ 
\multicolumn{1}{c}{$\varepsilon_{t=0}$} & \multicolumn{1}{c}{1}              \\ 
\multicolumn{1}{c}{$\gamma$}                      & \multicolumn{1}{c}{0.999}           \\ 
\multicolumn{1}{c}{$\omega_{max}$}                      & \multicolumn{1}{c}{4}              \\ \bottomrule
\end{tabular}
}
\captionof{table}{Default algorithm parameter settings.}
\end{minipage}
\begin{minipage}[c]{0.8\textwidth}
\centering
\scalebox{0.75}{
\begin{tabular}{cccc}
\toprule
$\omega$ = 1                                                        & $\omega$ = 2                                                                   & $\omega$ = 3                                                                    & $\omega$ = 4                                                                     \\ \midrule
\begin{tabular}[c]{@{}c@{}}1 searcher\\ ($\beta$ = 0)\end{tabular} & \begin{tabular}[c]{@{}c@{}}2 searchers\\ ($\beta$ = 0)\end{tabular}           & \begin{tabular}[c]{@{}c@{}}3 searchers\\ ($\beta$ =0)\end{tabular}              & \begin{tabular}[c]{@{}c@{}}4 searchers\\ ($\beta$ =0)\end{tabular}              \\ \addlinespace
\begin{tabular}[c]{@{}c@{}}1 shaper\\ ($\beta$ = 1)\end{tabular}   & \begin{tabular}[c]{@{}c@{}}1 searcher 1 shaper\\ ($\beta$ = 0.5)\end{tabular} & \begin{tabular}[c]{@{}c@{}}2 searchers 1 shaper\\ ($\beta$ = 0.33)\end{tabular} & \begin{tabular}[c]{@{}c@{}}3 searchers 1 shaper\\ ($\beta$ = 0.25)\end{tabular} \\ \addlinespace
-                                                             & \begin{tabular}[c]{@{}c@{}}2 shapers\\ ($\beta$ = 1)\end{tabular}             & \begin{tabular}[c]{@{}c@{}}1 searcher 2 shapers\\ ($\beta$ = 0.67)\end{tabular} & \begin{tabular}[c]{@{}c@{}}2 searchers 2 shapers\\ ($\beta$ = 0.5)\end{tabular} \\ \addlinespace
-                                                             & -                                                                        & \begin{tabular}[c]{@{}c@{}}3 shapers\\ ($\beta$  = 1)\end{tabular}              & \begin{tabular}[c]{@{}c@{}}1 searcher 3 shapers\\ ($\beta$ = 0.75)\end{tabular} \\ \addlinespace
-                                                             & -                                                                        & -                                                                          & \begin{tabular}[c]{@{}c@{}}4 shapers\\ ($\beta$ = 1)\end{tabular}               \\\bottomrule
\end{tabular}
\label{groupsC}
}
\captionof{table}{StructC group combinations.}
\end{minipage}

\subsection{Experimental results}~\label{results}
We investigate the performance of various models as a function of the problem complexity. We achieve this by visualizing and analysing the model behaviors during the search process. 
All models were validated using 50 runs with each run using a randomly generated fitness landscapes (same set of landscapes were used for each model) and lasting for 100 iterations. For StructC, group compositions were randomly sampled to preserve generality. As a result, it was ensured that each unique group composition will have appeared 50 times throughout the whole experiment. Each run takes around $2$ minutes depending on problem complexity using an Intel i7 (8th gen.) CPU, 8 GB DDR3L RAM.\footnote{The code to replicate these experiment can be downloaded at \url{https://github.com/BrandonWoei/NK-Landscape-Extensions}} 

\begin{figure*}[t]
\centering
\includegraphics[scale=0.32]{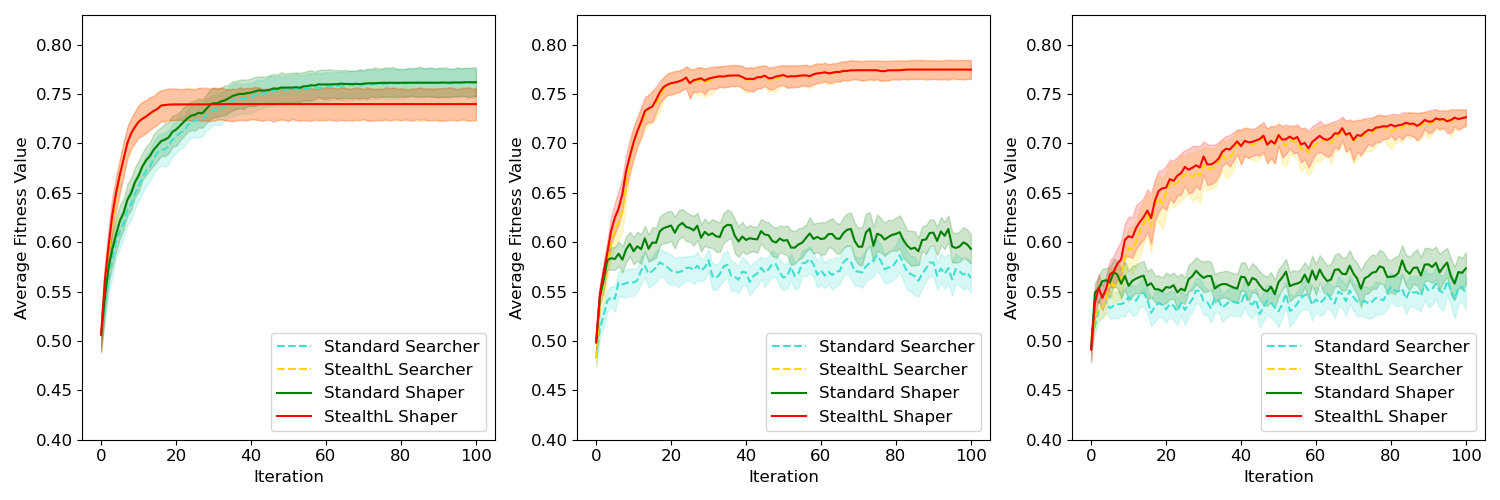}
\caption{Average best fitness (and 95\% confidence interval in shading calculated using $t$ statistic) obtained by the Standard Model and StealthL at (left plot) minimal ($K=0$, $E=0$), (middle plot) intermediate ($K=5$, $E=6$), and (right plot) high ($K=11$, $E=12$) landscape ruggedness and malleability. StealthL Searcher and Stealth Shaper have almost identical performance. }
\label{Figure1}
\end{figure*}

\begin{figure*}[t]
\centering
\includegraphics[scale=0.29]{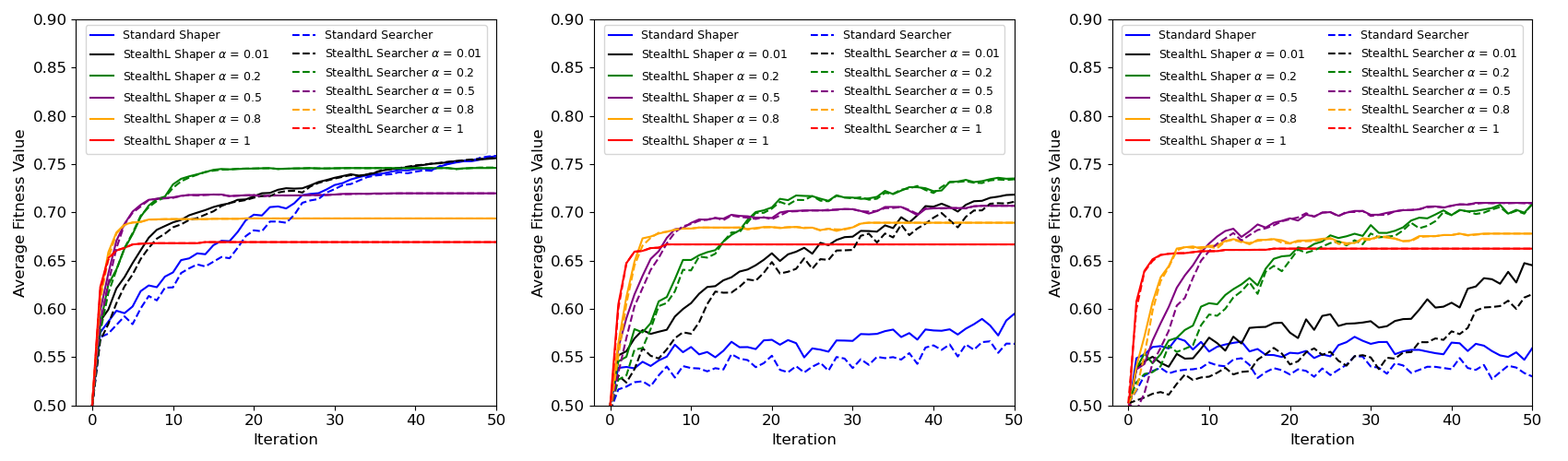}
\caption{Varying StealthL learning rate $\alpha$ at extreme malleability ($E=12$) with (left plot) minimal ($K=0$) (middle plot) intermediate ($K=5$) (right plot) high ($K=12$) landscape ruggedness. The confidence interval of the average fitness value has been omitted in this and following plots for the sake of readability.}
\label{Figure2}
\end{figure*}

\begin{figure*}[t]
\centering
\includegraphics[scale=0.285]{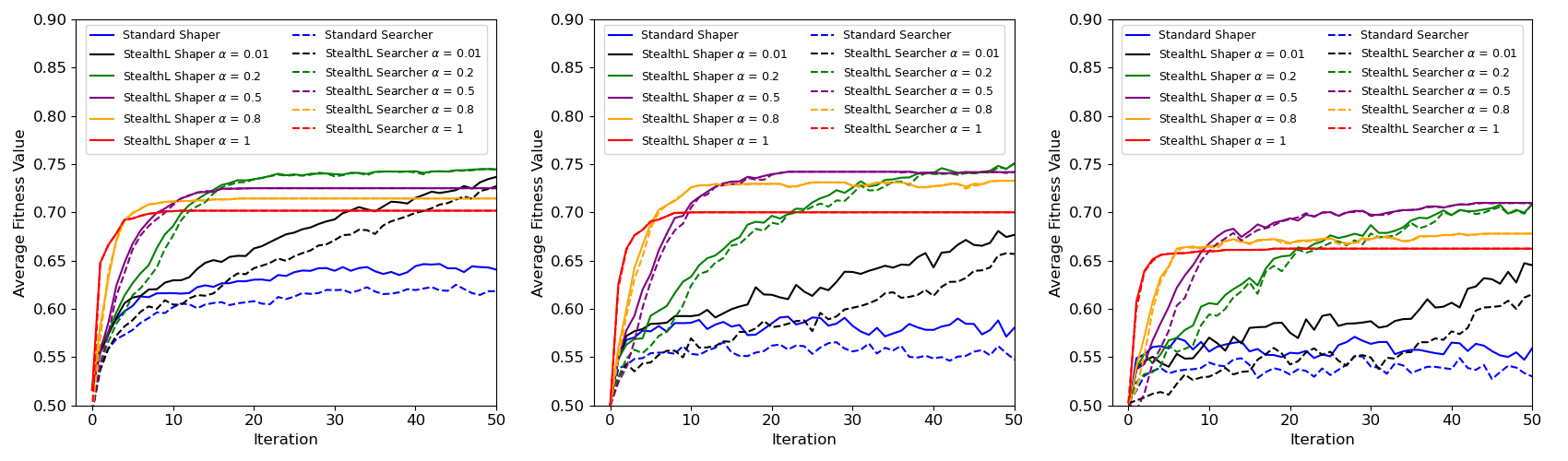}
\caption{Varying StealthL learning rate $\alpha$ at extreme ruggedness ($K=11$) with (left plot) minimal ($E=0$) (middle plot) intermediate ($E=6$) (right plot) high ($E=12$) landscape malleability.}
\label{Figure3}
\end{figure*}

\vspace{+1mm}
\noindent \textbf{Standard vs. Stealthy Global Learning\quad}Figures~\ref{Figure1}-~\ref{Figure3} compare the performance of searchers and shapers for the Standard Model and StealthL as well as for different learning rates ($\alpha$). Following observations can be made: 

Shapers and searchers in StealthL outperform their counterparts in the Standard Model significantly for rugged landscape ($K>0$) and regardless of the learning rate $\alpha$. However, for $K=0$, the Standard Model achieved a better performance because StealthL is suffering from premature convergence caused by its weakened random perturbation, a trade-off of guided learning. Stronger mutation is necessary when peaks on the fitness landscape are rare (or a single peak is present as is the case for $K=0$) and sufficiently far from one another since StealthL becomes complacent to the point in which exploration is inhibited when its corresponding agents all have roughly good and near solutions.
    
For StealthL, the performance of searchers and shapers is almost identical regardless of the level of ruggedness $K$. For the Standard Model, the performance gap between searchers and shapers depends on $K$ with the most significant performance gap being observed for an intermediate level of landscape ruggedness. 
    
The learning rate $\alpha$ has a significant impact on search performance. A decrease in $\alpha$ leads to a slower convergence but an improved final performance (if given sufficient optimization time). A high learning rate of $\alpha > 0.8$ leads to premature convergence. Generally, slightly higher learning rates perform better as the level of landscape ruggedness and malleability increases. Searchers and shapers are affected in a similar fashion by a changing learning rate, while the performance gap between searchers and shapers for a specific learning rate is minimal.


\begin{figure*}[t!]
\centering
\includegraphics[scale=0.32]{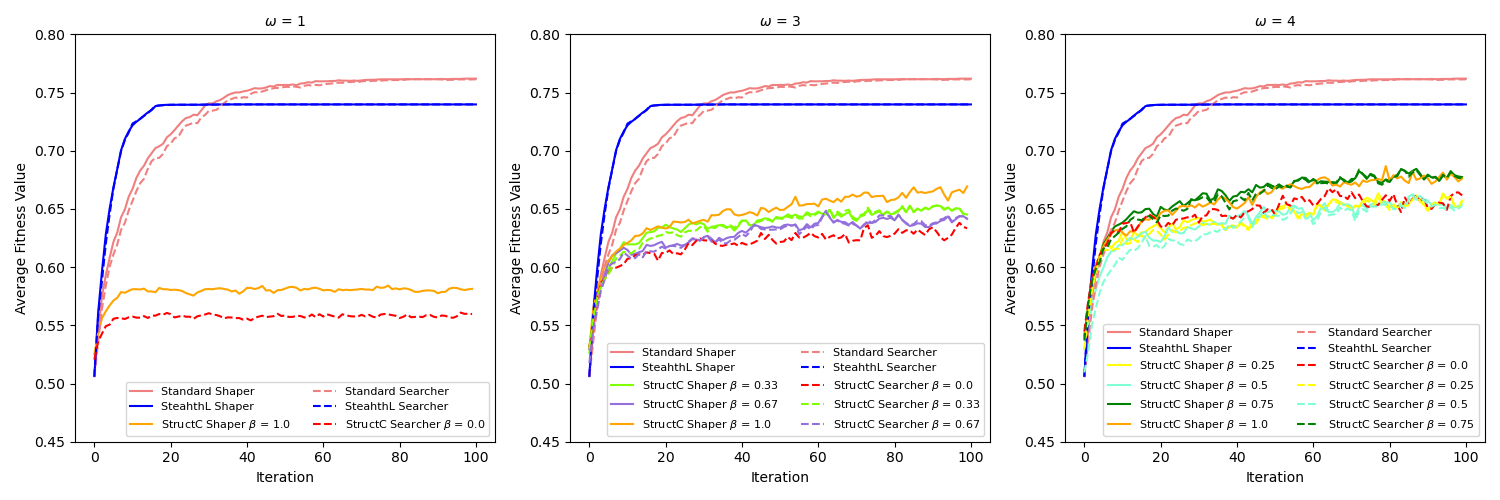}
\caption{Average fitness values achieved by StructC for different group sizes at low ruggedness and malleability ($K=0$, $E=0$).}
\label{Figure4}
\end{figure*}

\begin{figure*}[t!]
\centering
\includegraphics[scale=0.32]{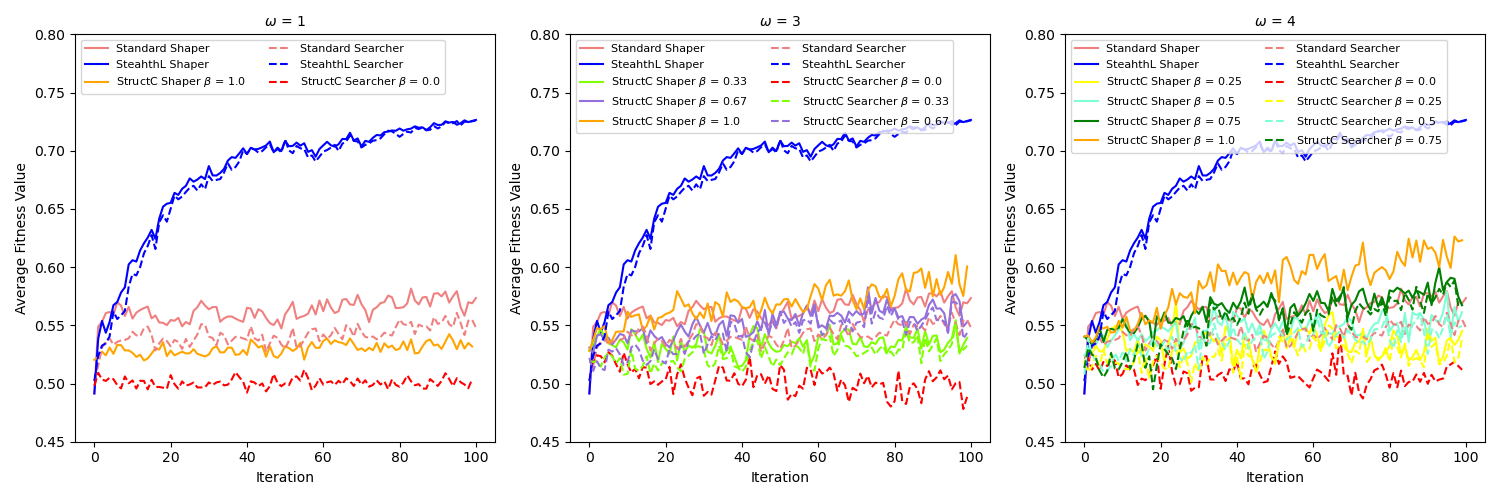}
\caption{Average fitness values achieved by StructC for different group sizes at high ruggedness and malleability ($K=11$, $E=12$).}
\label{Figure5}
\end{figure*}

\subsection{StructC Model at Low Landscape Ruggedness and Malleability}
Figure~\ref{Figure4} visualizes the impact of different group sizes on structured cooperation for $K=0$, $E=0$. Evidently, being in a bigger group with shapers helped with a searcher's performance. Lone searchers suffered from severe premature convergence with no sign of any improvement. The advantage of having a shaper was especially prominent in a twin group. However, the relationship between shaper proportion and searcher performance was highly inconspicuous and nonlinear. A group of size 3 gave the best searcher performance when it only had 1 shaper. The further increase in shapers decreased searcher performance but was still better than the complete searcher group. The results for size 4 groups were even more astonishing, with searchers only groups outperforming groups with 1 or 2 shapers in terms of searcher performance. However, a group with 3 shapers enabled its only searcher to outperform complete searcher groups by a noticeable degree. Whilst shapers are useful to searchers, the benefits were found to be highly dependent on group size and the number of shapers.

Albeit outperforming a lone searcher, a lone shaper suffers a similar premature convergence problem. For twin groups, a complete shaper group gave better shaper performance than a mixed group. For size 3 groups, a complete shaper gave the best shaper performance. In terms of the mixed groups, groups with 1 searcher were found give better shaper performance than groups with 2 searchers. For size 4 groups, shaper dominated groups gave better shaper performance than the balanced and searcher dominated groups. Unexpectedly, a shaper dominated group with 1 searcher gives slightly better shaper performance than a complete shaper group. Generally, a large group size dominated by shapers ($>75\%$) was found to be group configuration that produced excellent searchers and shapers. Predictably, all groups were not able to surpass the standard and StealthL model performance under all conditions.

\subsection{StructC Model at High landscape Ruggedness and Malleability}

Figure~\ref{Figure5} visualizes the impact of different group sizes on structured cooperation for $K=11$, $E=12$. At high ruggedness and malleability, lone searchers had no improvement from its initial fitness. Mixed twin groups were only able to marginally improve their initial searcher performance. Searcher performance of searcher dominated twin groups was seen to deteriorate slightly with time. Such temporal deterioration effects were also seen in all complete searcher groups regardless of group size. Universally, the increase in group size and shaper proportion improved searcher performance. At maximum group size, searcher performance was able to exceed that of the standard model by a noticeable degree. On the other hand, lone shapers found it difficult to improve themselves. Similarly, a larger group size with more shapers led to excellent shaper performance.

\section{Conclusion}\label{conclusion}
We have used a multi-agent system to model how agents (firms) may collaborate and adapt in a business `landscape' where some, more influential, firms are given the power to \emph{shape}\/ the landscape of other firms. We have found that shapers outperform searchers under all landscape conditions. However, excessive landscape reshaping can lead to poor collective performance due to the instability it introduces. Additionally, both searchers and shapers perform best, even under dynamic business landscapes, when they can keep their organisational complexities at a minimum by reducing relationships between elements of their policies (encoded here by the two parameters $K$ and $E$). Complex organisations can find it hard to cope with landscape changes especially when the changes are frequent and substantial. However, we found that this can be overcome to some extent  by allowing for collaboration via experience-sharing between firms. Whilst mutual learning is beneficial, direct mimicry of best practices can lead to a reduction in collective knowledge as firm' shared inertia hinders exploration, thereby weakening the synergistic effects of collaboration. Lastly, the positive effects of collaboration were also found to be at their best when collaboration groups are dominated by shapers. 

Despite having extended NK models, a limitation of our work is that reality is more complex than the abstraction considered here, particularly in terms of the kinds of strategy different firms might employ. Moreover, the model could be made more realistic by, for example, accounting for additional objectives and constraints, observing delays between evaluating an objective/constraint function and having its value available~\cite{allmendinger2015multiobjective,chugh2018surrogate,wang2021transfer}, and simulating more than two types of agents as firms may vary widely e.g. in terms of capabilities, goals, and how they engage with other firms.

\bibliographystyle{splncs04}






\end{document}